\documentclass{jfm}
\usepackage{natbib, epsfig, amsmath, amssymb, tikz, graphicx}
\usepackage{mathrsfs, bm}
\usepackage{color}
\usepackage{epstopdf}
\usepackage{float}
\graphicspath{{../fig/}}
\usepackage[colorlinks=true,linkcolor=blue,citecolor=blue]{hyperref}

\def \at {a_{t}}

\def \Rey  {\mbox{Re}}

\def \At  {\mbox{At}}
\def \Bo  {\mbox{Bo}}

\usepackage[normalem]{ulem}
\newcommand{\beq}{\begin{equation}}
\newcommand{\eeq}{\end{equation}}

\newcommand{\REM}[1]{{}}

\newcommand{\Eq}[1]{Eq.~(\ref{#1})}
\newcommand{\fig}[1]{figure~\ref{#1}}
\newcommand{\subfig}[2]{figure~\ref{#1}(#2)}

\shorttitle{Energy spectra in 3D buoyancy driven bubbly flows}
\shortauthor{Vikash Pandey, Rashmi Ramadugu, Prasad Perlekar}

\begin{document}

\title{Liquid velocity fluctuations and energy spectra in three-dimensional buoyancy driven bubbly flows}
\author{Vikash Pandey \aff{1},
Rashmi Ramadugu \aff{1}
\and Prasad Perlekar \aff{1}
\corresp{\email{perlekar@tifrh.res.in}}} 

\affiliation{\aff{1} TIFR Center for Interdisciplinary Sciences, Hyderabad, 500107}

\maketitle

\begin{abstract}
We present a direct numerical simulation (DNS) study of pseudo-turbulence in buoyancy driven bubbly flows for 
a range of Reynolds ($\Rey$) and Atwood ($\At$) numbers. We study the probability
distribution function of the  horizontal and vertical liquid velocity fluctuations and
find them to be in quantitative agreement with the experiments.  The energy spectrum shows
the $k^{-3}$ scaling at high $\Rey$ and becomes steeper on reducing the $\Rey$.  To
investigate the spectral transfers in the flow, we derive the scale-by-scale energy budget
equation.  Our analysis shows that, for scales smaller than the bubble diameter, the net
production because of the surface tension and  the kinetic energy flux  balances viscous
dissipation  to give the $k^{-3}$ scaling of the energy spectrum for both low and high
$\At$.
\end{abstract}

\section{Introduction}
Bubble laden flow appears in a variety of natural \citep{clift1978,gonnermann2007fluid} and
industrial \citep{deckwer} processes. Presence of bubbles dramatically alters the transport
properties of a flow \citep{mudde_rev_2005,cecc10,RBLuca,pan17,riss18,elgh19,mat19}.  A
single bubble of diameter $d$, because of
buoyancy, rises under gravity. Its trajectory and the wake flow depend on the density and
viscosity contrast with the ambient fluid, and the surface tension \citep{clift1978,
bha81,tri15}.  A suspension of such bubbles at moderate volume
fractions generates complex spatiotemporal flow patterns that are often referred to as
pseudo-turbulence or bubble-induced agitation \citep{mudde_rev_2005,riss18}.

Experiments have made significant progress in characterizing velocity fluctuations of the fluid
phase in pseudo-turbulence. A key observation is the robust power-law scaling in the energy spectrum with an exponent of
$-3$ either in frequency $f$ or the wave-number $k$ space
\citep{martinez_2010,risso_legendre_2010,mendez}. The scaling range, however,
remains controversial. \citet{risso_legendre_2010} investigated turbulence in the wake of a
bubble swarm and found the $k^{-3}$ scaling for length scales larger than the bubble diameter
$d$ (i.e., $k<2\pi/d$), whereas \citet{martinez_2010,vivek_2016} observed this scaling for
scales smaller than $d$ in a steady state bubble suspension. Experiments on bouyancy driven bubbly 
flows in presence of grid-turbulence \citep{lance_1991,vivek_2016,almeras2017} observe
Kolmogorov scaling for scales larger than the
bubble diameter and smaller than the forcing scale and a much steeper $k^{-3}$ scaling for scales 
smaller than the bubble diameter and larger than the dissipation scale. \citet{lance_1991}
argued that, assuming production because of wakes to be local in spectral space, balance
of  production with viscous dissipation leads to the observed  $k^{-3}$ scaling.

Fully resolved numerical simulations of three-dimensional (3D) bubbly flows  for a range of
Reynolds number $O(10) < \Rey < O(10^3)$ \citep{roghair,bunner_tryg_2002,bunner_rvel_2002} found the $k^{-3}$ scaling for
length scales smaller than $d$ ($k>2\pi/d$) and attributed it to the balance between viscous dissipation
 and the energy production by the wakes \citep{lance_1991}.

Two mechanisms proposed to explain the observed scaling behavior in experiments are: $(i)$
superposition of velocity fluctuations generated in the vicinity of the bubbles
\citep{rissou_2011}, and $(ii)$ at high $\Rey$, the instabilities in the flow through
bubble swarm  \citep{lance_1991,mudde_rev_2005,riss18}. In an experiment or a simulation, it is difficult 
to disentangle these two mechanisms.

In classical turbulence, a constant flux of energy is maintained between the injection and
dissipation scales \citep{frisch,per09,boffetta}. In pseudo-turbulence, on the other hand,
it is not clear how the energy injected because of buoyancy is transferred between
different scales. In particular, the following key questions remain unanswered: $(i)$ How do  liquid 
velocity fluctuations and the pseudo-turbulence spectrum depend on the Reynolds number ($\Rey$)? $(ii)$ What is the energy 
budget and the dominant balances? $(iii)$ Is  there an energy cascade (a non-zero energy flux)?

In this paper, we address all of the above questions for experimentally relevant Reynolds ($\Rey$) and Atwood ($\At$) numbers. 
We first investigate the dynamics of an isolated bubble and show that the wake flow behind the bubble is in 
agreement with earlier experiments and simulations. Next for a bubbly suspension we show
that the the liquid velocity fluctuations are in quantitative agreement with the
experiments of \citet{risso_legendre_2010} and the bubble velocity
fluctuations are in quantitative agreement with the simulations of
\citet{roghair}. We then proceed to derive the scale-by-scale energy budget equation and
investigate the dominant balances for different $\Rey$ and $\At$. We find that  for scales
smaller than the bubble diameter, viscous dissipation balances net nonlinear transfer of
energy because of advection and the surface tension  to give $k^{-3}$ psuedo-turbulence
spectrum. Intriguingly, the dominant balances are robust and do not depend on the density
contrast ($\At$).

\section{Model and Numerical Details}
We study the dynamics of bubbly flow by using  Navier-Stokes (NS) equations with a surface
tension force because of bubbles

\begin{subequations}
	\begin{eqnarray}
	\rho D_t\bm{u} &=&   \nabla \bm{\cdot} [2 \mu {\cal S}]  - \nabla p 
	 + {\bm F}^\sigma  + {\bm F}^g, \label{eqn:mom}\\ 
	  \nabla\bm{\cdot}\bm{u} &=& 0.
	\end{eqnarray}
	\label{eqn:ns}
\end{subequations}

Here, $D_t = \partial_t + (\bm{u}\cdot\nabla)$ is the material derivative, ${\bm u} = (u_x,u_y,u_z)$ is the hydrodynamic
velocity, $p$ is the pressure, ${\cal
 S}\equiv (\nabla {\bm u} + \nabla {\bm u}^T)/2$ is the rate of deformation tensor,  $\rho
\equiv \rho_f c + \rho_b (1-c)$ is the density, $\mu\equiv \mu_f c + \mu_b (1-c)$ is the
viscosity, $\rho_f$ ($\rho_b$)  is the fluid (bubble) density, and  $\mu_b$ ($\mu_f$) is
the bubble (fluid) viscosity.  The value of the indicator function $c$ is equal to zero in
the bubble phase and  unity in the fluid phase. The surface tension force is
$\bm{F}^\sigma \equiv \sigma \kappa \hat{\bm{n}}$, where $\sigma$ is the coefficient of
surface tension,
$\kappa$ is the curvature, and $\hat{\bm{n}}$ is the normal to the bubble interface. ${\bm F}^{g} \equiv [\rho_a-\rho]
g\hat{\bm{z}}$ is the buoyancy force, where $g$ is the accelaration due to gravity, and
$\rho_a \equiv  [\int \rho(c) d{\bm x}]/L^3$ is the average density. For small Atwood
numbers, we  employ Boussinesq approximation whereby, $\rho$ in the left-hand-side of
\Eq{eqn:mom} is replaced by the average density $\rho_a$.
 
We solve the Boussinesq approximated NS using a pseudo-spectral method \citep{canuto}
coupled to a front-tracking algorithm \citep{Tryg2001, paris} for bubble dynamics. Time
marching is done using a scond-order Adams-Bashforth scheme. For the non-Boussinesq NS, we
use the open source finite-volume-front-tracking solver PARIS \citep{paris}.

We use a cubic periodic box of volume $L^3$ and discretize it with $N^3$ collocation points.  We
initialize the velocity field ${\bm u}=0$ and place the centers of $N_b$ bubbles at random
locations such that no two bubbles overlap. The Reynolds number $\Rey$, the Bond number
$\Bo$, and the bubble volume fraction $\phi \equiv [\int (1-c) d{\bm x}]/L^3$ that we use
(see table~\ref{tab:runs}) are comparable to the experiments \citep{mendez, risso_legendre_2010}.

 \begin{table}
	\begin{center}
          \begin{tabular}{lcccccccccccccc}
             {\tt runs}   & $L$  & $N$   & $N_b$  &$d$  & $g$& $\mu_f$ & $\phi\%$   & $\Rey$& $\At$ &
                     $\Bo$   & $\epsilon_\mu$  & $\epsilon_w$ & $\epsilon_{\mu,f}$& $\epsilon_{inj}$ \\

             $\tt{R1}$    & $256$ & $512$ &  $60$ & $24$  & $1.0$& $0.32$ & $2.6$ & $104$ & $0.04$  & $1.8$
               &$3.6\cdot10^{-3}$ & $2.8\cdot 10^{-3}$ &$2.0\cdot 10^{-3}$& $3.5\cdot10^{-3}$   \\

             $\tt{R2}$    & $256$ & $512$  &  $60$ & $24$ & $1.0$ & $0.20$ & $2.6$ & $166$ & $0.04$  & $1.0$
              &$4.3\cdot10^{-3}$ & $2.8\cdot 10^{-3}$ &$2.6\cdot 10^{-3}$& $4.3\cdot10^{-3}$ \\

             $\tt{R3}$    & $128$ & $432$ &  $10$ & $22$ & $8.75$ & $0.42$  & $2.6$  &   $206$ & $0.04$  & $2.1$
               &$9.5\cdot10^{-2}$ &$7.1\cdot 10^{-2}$& $6.7\cdot 10^{-2}$& $9.4\cdot10^{-2}$   \\

            $\tt{R4}$    & $128$ & $432$ &  $10$ & $22$  & $10.5$ & $0.32$ & $2.6$ &   $296$  & 0.04  & $1.9$
                & $1.3\cdot10^{-1}$ & $9.4\cdot 10^{-2}$ &$9.6\cdot 10^{-2}$& $1.3\cdot10^{-1}$ \\
                  
            $\tt{R5}$    & $256$ & $256$ & $40$ & $24$ & $0.1$ & $0.32$  & $1.7$ &   $113$ & $0.90$  & $2.0$ 
                 &$3.2\cdot10^{-3}$ & $2.4 \cdot 10^{-3}$ & $1.8 \cdot 10^{-3}$& $3.0\cdot10^{-3}$ \\
            $\tt{R6}$    & $256$ & $256$ & $40$ & $24$ & $1.0$ & $0.32$  & $1.7$ &   $345$ & $0.80$  & $2.4$ 
                 &$8.1\cdot10^{-2}$ & $6.9 \cdot 10^{-2}$ & $5.4 \cdot 10^{-2}$& $8.4\cdot10^{-2}$ \\

            $\tt{R7}$    & $256$ & $256$ &  $40$ & $24$ & $1.0$ & $0.32$ & $1.7$ &  $358$ & $0.90$  & $1.9$ 
                 &$1.0\cdot10^{-1}$ & $7.7 \cdot 10^{-2}$ & $6.2 \cdot 10^{-2}$& $1.0\cdot10^{-1}$  \\
                  
    \end{tabular}

 \caption{\label{tab:runs} Table of parameters used in our DNS. Here, $\delta \rho\equiv \rho_f-\rho_b$ is the density difference,  
 $\Rey \equiv \sqrt{\rho_f \delta \rho g d^{3}}/\mu_f$ is the Reynolds number, $\Bo \equiv
 \delta\rho gd^2/\sigma$ is the Bond number,  $\At = \delta \rho/(\rho_f + \rho_b)$ is the
 Atwood number,  $\epsilon_w \equiv \phi(\delta \rho g d /\rho_f)^{3/2}/d$ is the estimate
 of the
     energy dissipation rate because of the bubble wakes (\cite{lance_1991}), and $\epsilon_{\mu,f}$ is  the 
     energy dissipation rate in the fluid phase.  We fix the value of the fluid density $\rho_f=1$ and assume
     same viscosity for the fluid and the bubble phase $\mu_f/\mu_b=1$.} 
\end{center}
\end{table}

\section{Results}
 
In subsequent sections, we investigate statistical properties of stationary
pseudo-turbulence generated in buoyancy driven bubbly flows.
Table~\ref{tab:runs} lists the parameters used in our simulations. Our parameters are
chosen such that the Reynolds number, Bond number, and the volume fraction are comparable
to those used in earlier experiments \citep{risso_legendre_2010,mendez,riss18}. We
conduct simulations at both low and high-$\At$ numbers to investigate role of density
difference on the statistics of pseudo-turbulence.
The rest of the paper is organized as follows. In \S\S~\ref{sbub} we study the trajectory
of an isolated bubbles and, consistent with the experiments, show that the bubble shape is
ellipsoidal. In \S\S~\ref{ekin}-\ref{pvel} we investigate the total kinetic energy budget
and the fluid and bubble centre-of-mass velocity fluctuations and make quantitative
comparison with the experiments.  Finally, in \S\S~\ref{esp} we study the kinetic energy
spectrum and scale-by-scale energy budget analysis. We present our conclusions in
\S~\ref{concl}.

\subsection{Single bubble dynamics}
\label{sbub}
In this section we study the dynamics an initially spherical bubble as it rises because of
buoyancy.  The seminal work of \citet{bha81} characterized the shape and trajectory of an isolated bubble 
in terms of Reynolds and Bond number.  Experiments on turbulent bubbly flows
\citep{lance_1991, vivek_2016} observe ellipsoidal bubbles. In the following, we
characterize the dynamics of an isolated bubble for
the parameters used in our simulations. 

\begin{figure}
    \includegraphics[width=0.32\linewidth]{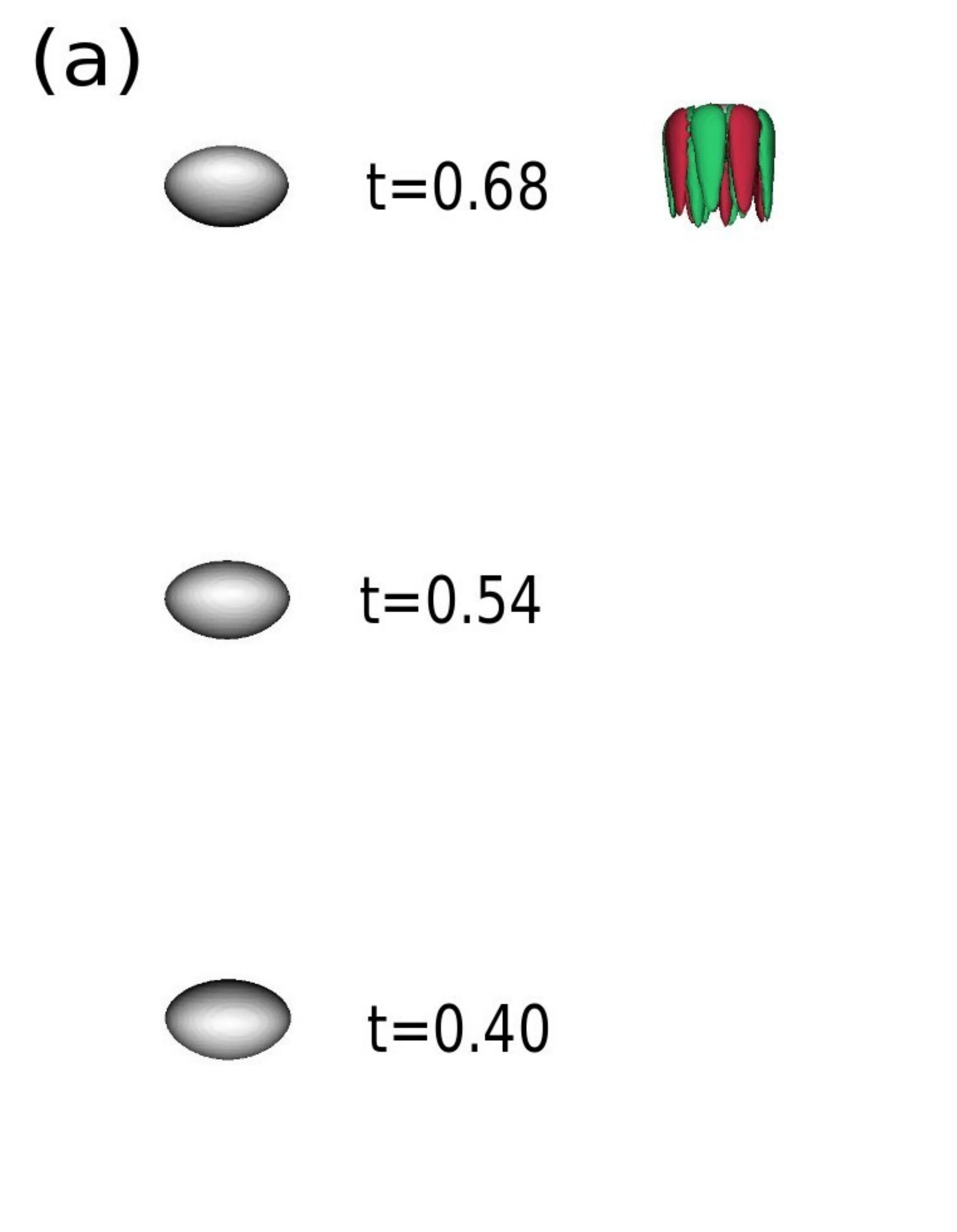}
    \includegraphics[width=0.32\linewidth]{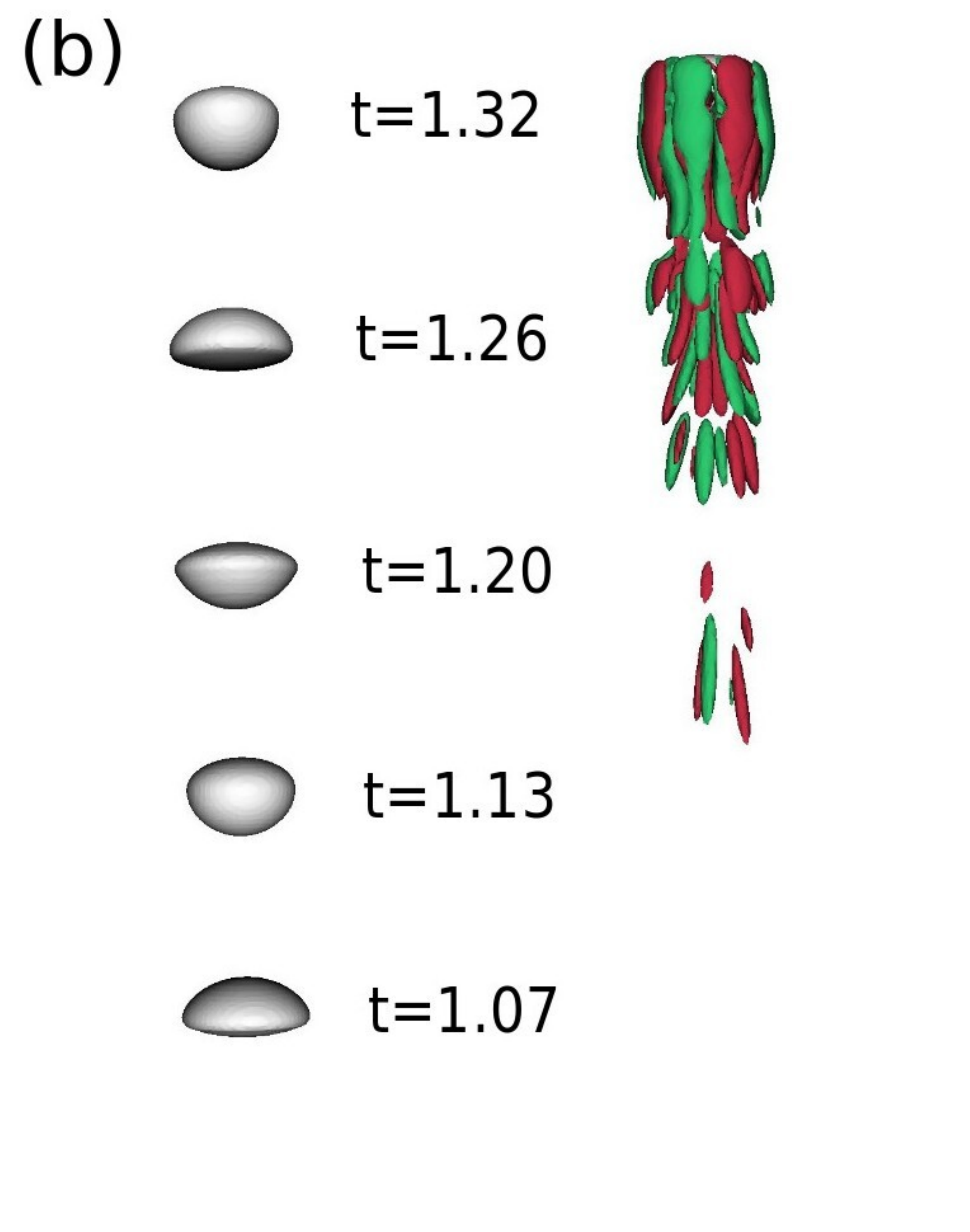}
    \includegraphics[width=0.32\linewidth]{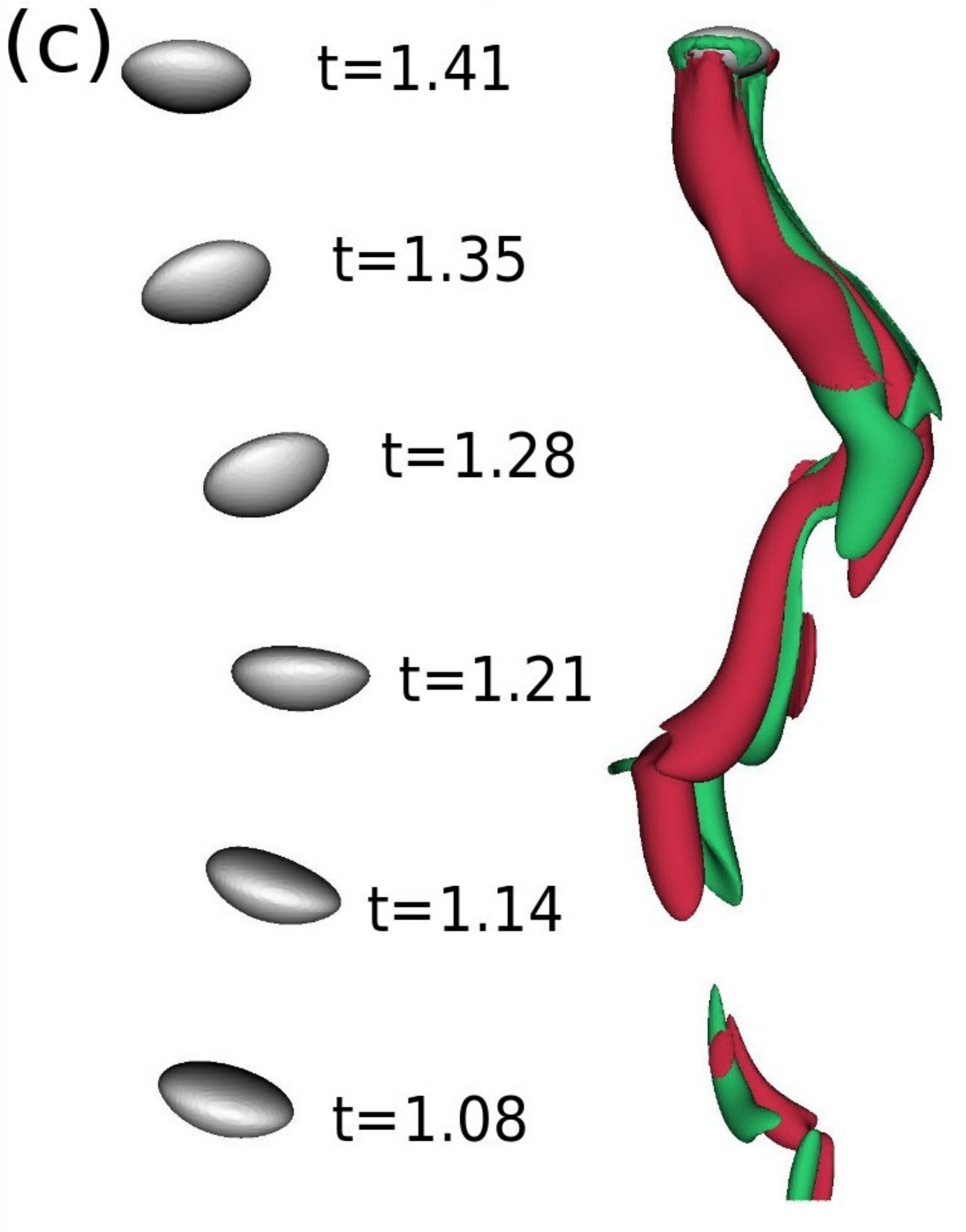}
   \caption{\label{sing:bub} Bubble positions at different times (in units
      of $\tau_s$) and the z-component of
      the vorticity ($\omega_z =  \partial_xu_y - \partial_yu_x$) for the case of single
      bubble rising under gravity. The non-dimensional parameters in representative cases are taken same as
      run {\tt R1} in panel (a), run {\tt R4} in panel  
      (b), and run {\tt R6} in panel (c). Green region corresponds to $\omega_z < 0$, whereas red
    region corresponds to $\omega_z > 0$. We plot iso-contours corresponding to $|\omega_z| = \pm 10^{-3}$ in (a),
    $|\omega_z| = \pm 10^{-2}$ in (b), and $|\omega_z| = \pm 10^{-1}$ in (c).}
    \end{figure}

To avoid the interaction of the bubble with its own wake, we use a vertically
elongated cuboidal domain of dimension $5d \times 5d \times 21d$.  After the bubble  rise
velocity attains steady-state,  \subfig{sing:bub}{a-c} shows the bubble shape and the vertical
component of the vorticity $\omega_z = (\nabla \times {\bm u}) \cdot \hat{z}$. For $\Rey=104$
and $\At=0.04$ (run {\tt R1}), the bubble shape is oblate ellipsoid and it rises in a
rectilinear trajectory. On increasing the $\Rey=295$ (run {\tt R4}), the bubble pulsates while
rising and sheds varicose vortices similar to \citet{pivello_2014}. Finally, for high
$\At=0.80$ and $\Rey=345$ (run {\tt R6}), similar to region III of
\citet{tri15}, we find that the bubble shape is oblate ellipsoid and it follows a zigzag
trajectory.

\begin{figure}
    \centering
    \includegraphics[width=0.9\linewidth] {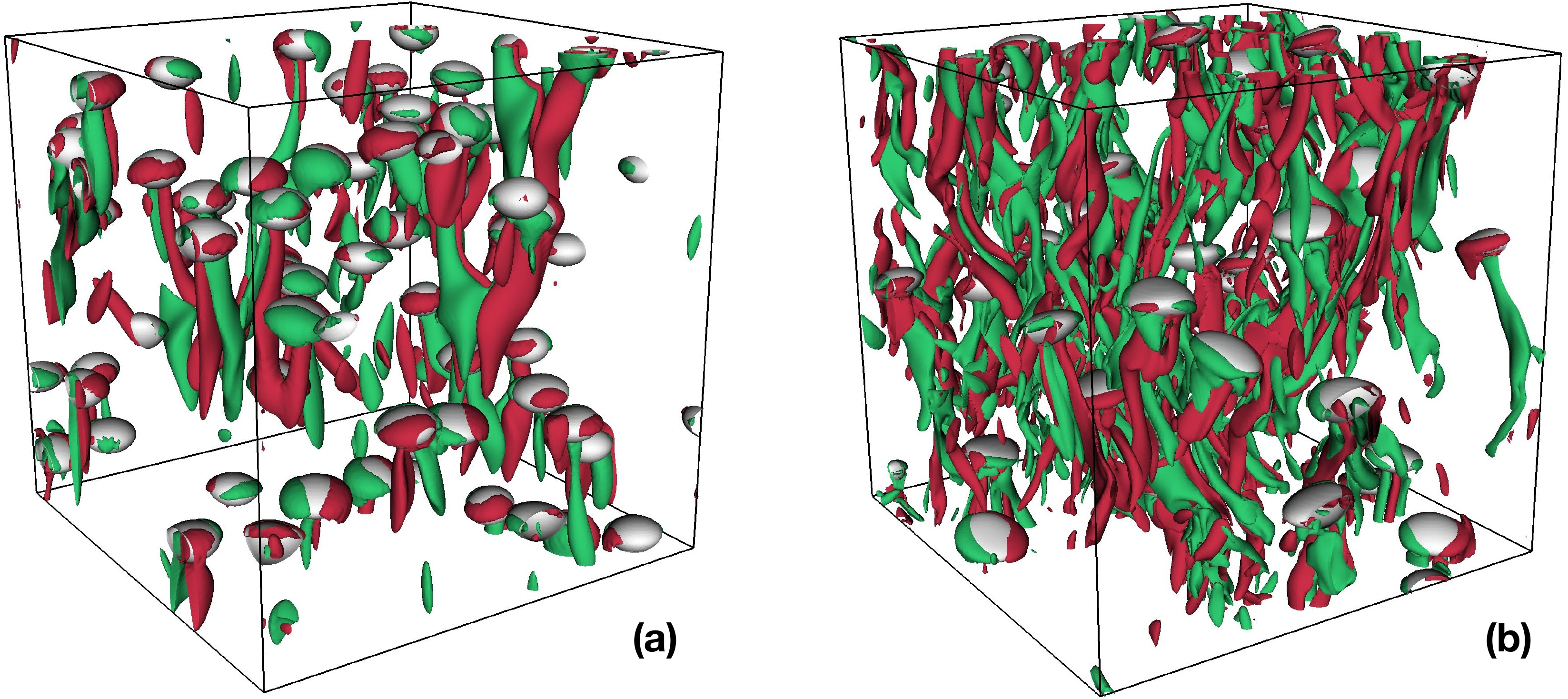}\\ 
    \includegraphics[width=0.45\linewidth]{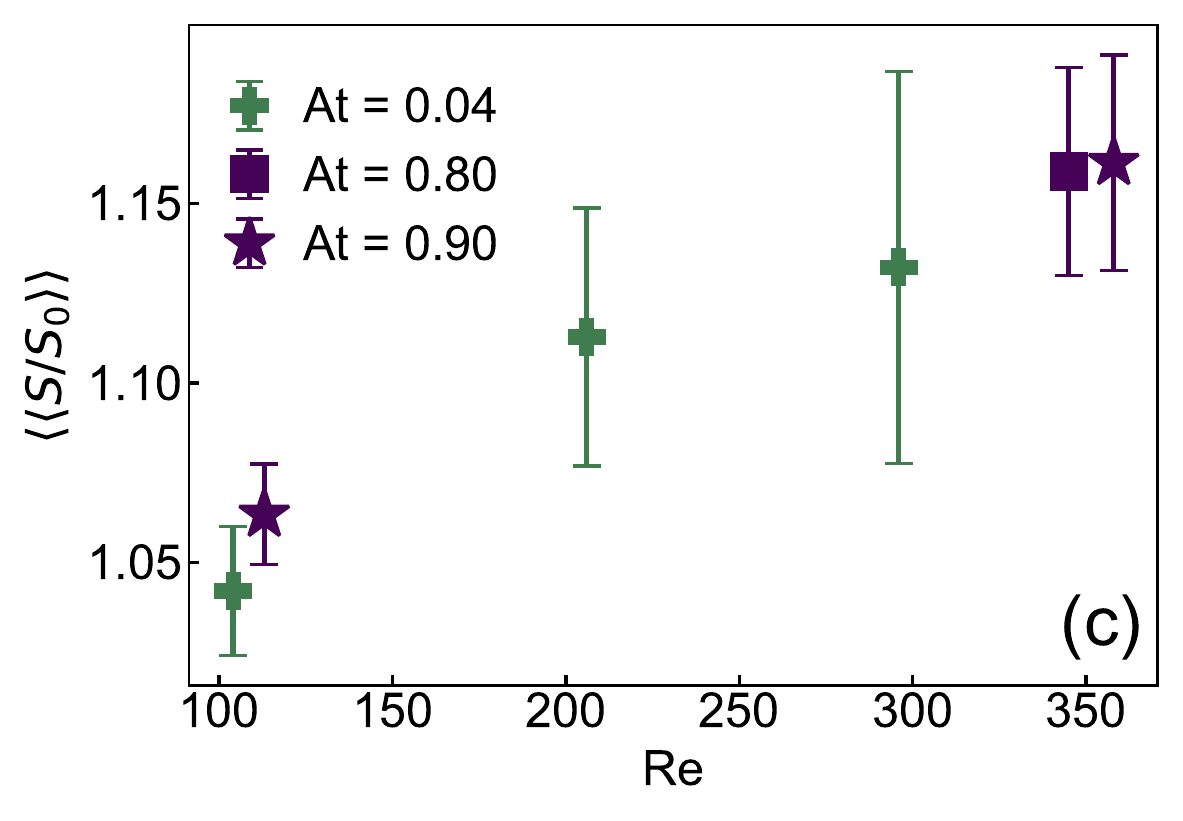} 
    \includegraphics[width=0.45\linewidth]{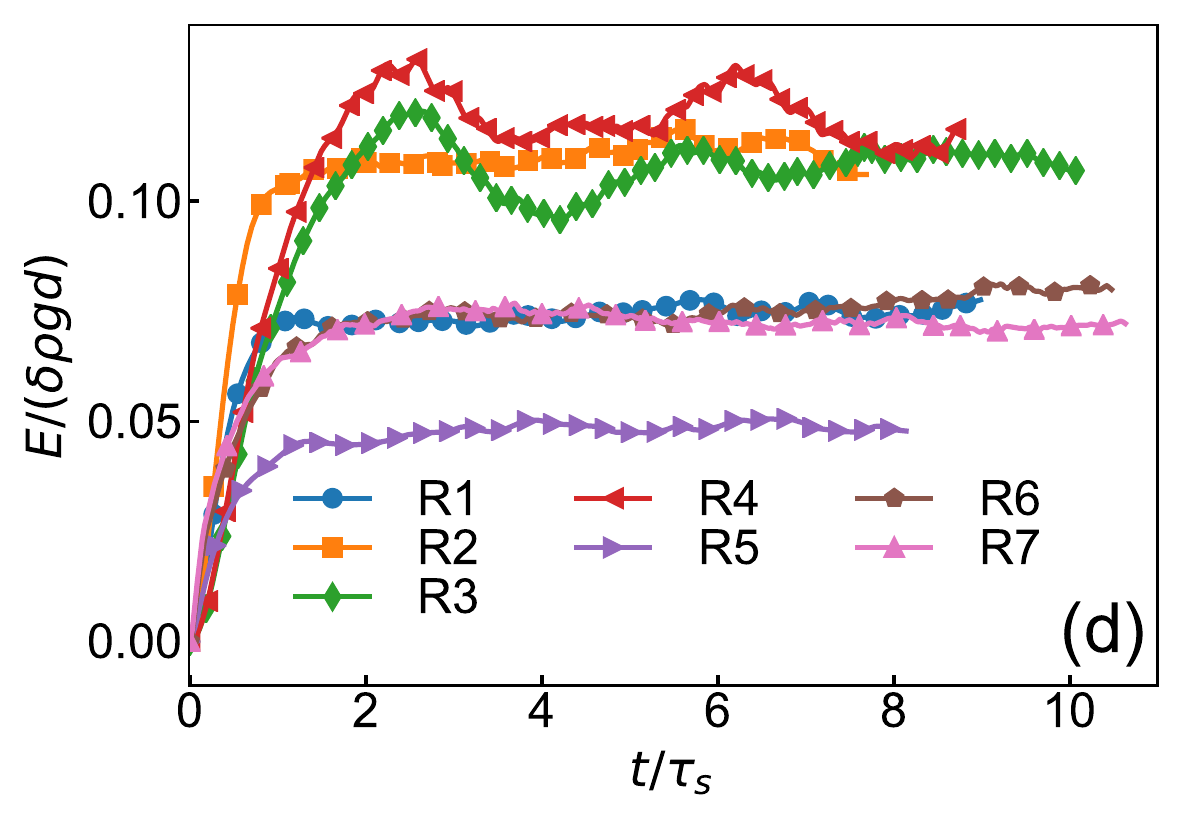} 
    \caption{\label{ener:bub}  Representative steady-state snapshot of the bubbles
      overlayed on the iso-contour plots of the $z$-component of the vorticity field
      $\omega_z\equiv [\nabla \times {\bm u}]\cdot \hat{\bm z}$ for $\Rey=104$, $\At=0.04$
      (a) and for $\Rey=345$, $\At=0.78$ (b). Regions with $\omega_z= 2\sigma_\omega (-2
      \sigma_\omega)$ are shown in red (green), where $\sigma_\omega$ is the standard
      deviation of $\omega_z$. As expected, bubble-wake interactions becomes more intense
      on increasing $\Rey$. (c) $\Rey$ versus average bubble deformation $\langle\langle
      S(t)/S(0)\rangle\rangle$ for low $\At=0.04$ and high $\At$ numbers, and (d) Kinetic
      energy evolution for the runs given in table~\ref{tab:runs}.}
\end{figure}

\subsection{Bubble suspension and kinetic energy budget}
\label{ekin}
The plots in \subfig{ener:bub}{a,b} show the representative steady state iso-vorticity
contours of the $z$-component of the vorticity along with the bubble interface position
for our bubbly flow configurations. As expected from our isolated bubble study in the
previous section, we observe rising ellipsoidal bubbles and their wakes which interact to
generate pseudo-turbulence. The individual bubbles in the suspension show shape
undulations which are similar to their isolated bubble counterparts [see movies available in the
supplementary material]. Furthermore, for  comparable $\Bo \approx 2$, the average bubble
deformation $\langle\langle S(t)/S(0) \rangle\rangle$ increases with increasing $\Rey$
[\subfig{ener:bub}{c}]. Here, $\langle \langle \cdot \rangle \rangle$ denote temporal
averaging over bubble trajectories in the statistically steady state, $S(t)$ is the
surface area of the bubble, and $S(0)=\pi d^2$.

The time evolution of the kinetic energy $E=\langle \rho u^2 /2 \rangle$ for runs {\tt R1
  - R7} is shown in \fig{ener:bub}(d).  A statistically steady state is attained around
$t\approx 2.5 \tau_s $, where $\tau_s=L/\sqrt{\delta \rho g d/\rho_f}$ is the approximate
time taken by an isolated  bubble to traverse the entire domain.  Using \Eq{eqn:mom}, we
obtain the total kinetic energy $E$ balance equation as

\begin{equation}
\partial_t \underbrace{\langle \frac{\rho {\bm u}^2}{2} \rangle}_{E}
 = -\underbrace{ 2 \langle \mu(c) \mathcal{S}:\mathcal{S} \rangle}_{\epsilon_\mu} 
  + \underbrace{\langle [\rho_a -\rho(c)] u_y g \rangle}_{\epsilon_{inj}} + 
  \underbrace{\langle \bm{F}^\sigma\cdot \bm{u} \rangle}_{\epsilon_{\sigma}}, 
\end{equation}
where, $\langle \cdot \rangle$ represents spatial averaging.
In steady state, the energy injected by buoyancy $\epsilon_{inj}$ is balanced by viscous
dissipation $\epsilon_{\mu}$. The energy injected by buoyancy $\epsilon_{inj} \approx
(\rho_f-\rho_b) \phi g \langle U\rangle$ where $\langle U \rangle$ is the average bubble
rise velocity. Note that $\epsilon_\sigma = -\partial_t\int\sigma ds$ \citep{joseph_1976}, 
where $ds$ is the bubble surface element, and its contribution is zero in the
steady-state.  The excellent agreement between
steady state values of $\epsilon_{\mu}$ and $\epsilon_{inj}$ is evident from table~\ref{tab:runs}. 

\citet{lance_1991} argued that the energy injected by the buoyancy is dissipated in the wakes 
on the bubble. The energy dissipation in the wakes can be estimated as $\epsilon_w = C_d
\phi ((\delta \rho/\rho_f)g d)^{3/2}/d$, where
$C_d$ is the drag coefficient. Assuming $C_d=O(1)$, we find that $\epsilon_w$ is indeed
comparable to the viscous dissipation in the fluid phase $\epsilon_{\mu,f}$ (see
table~\ref{tab:runs}).

\subsection{Probability distribution function of the fluid and bubble velocity fluctuations}
\label{pvel}
  
\begin{figure}
\begin{center}
    \includegraphics[width=0.48\linewidth]{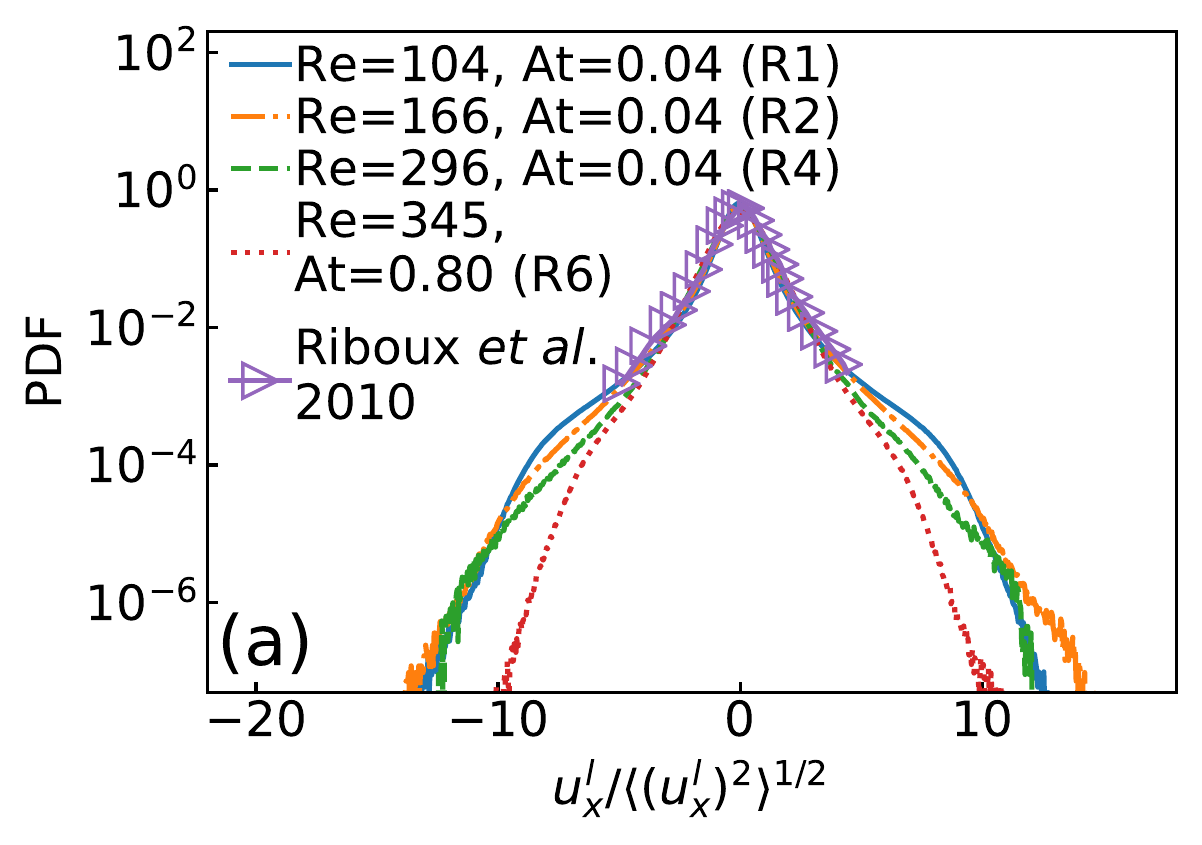}
    \includegraphics[width=0.48\linewidth]{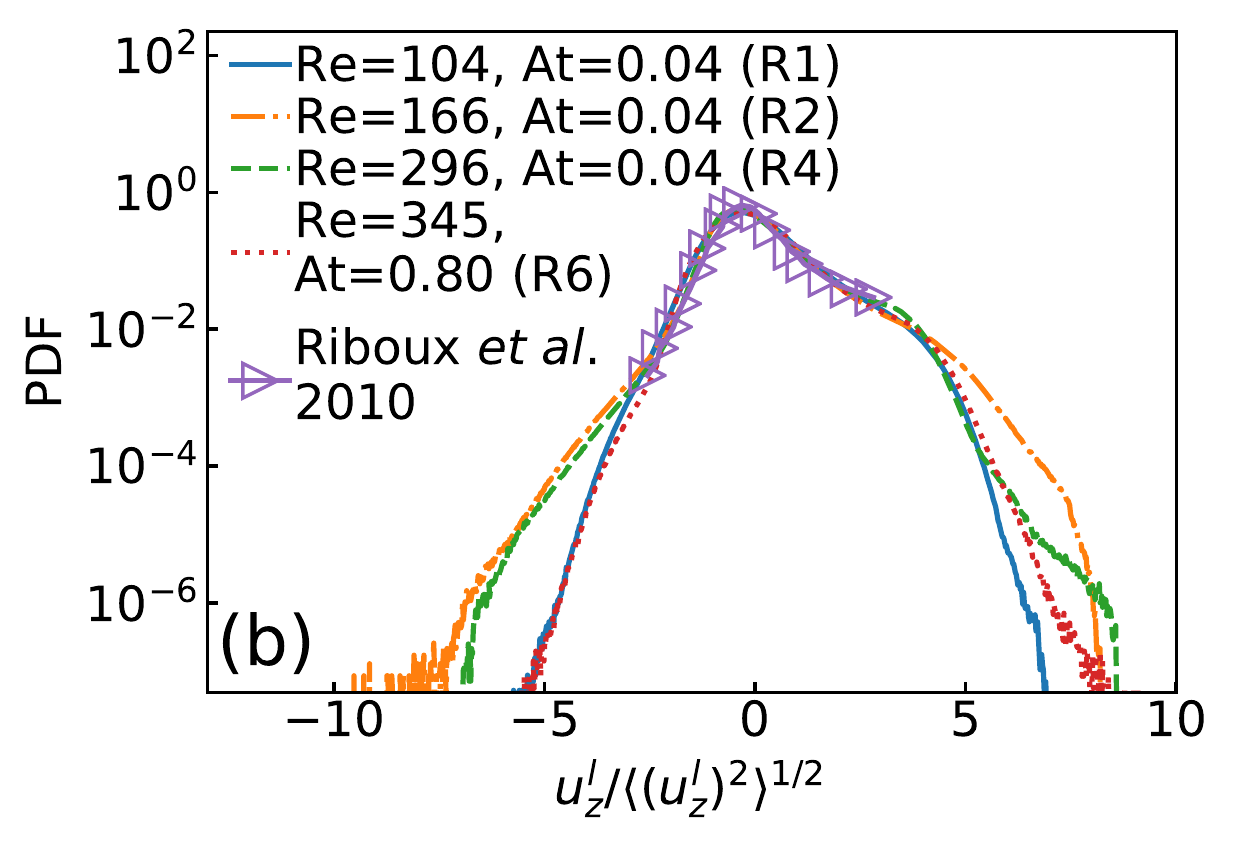}
    \end{center}
\caption{\label{velpdf3d} The probability distribution function of the (a) horizontal
    component (b) vertical component of the liquid velocity fluctuations for runs given in
    table \ref{tab:runs}. The p.d.f. obtained from our DNS are in excellent agreement with 
    the experimental data of \citet{risso_legendre_2010}
    [Data extracted using enguage https://markummitchell.github.io/engauge-digitizer/].}
\end{figure}

 In \subfig{velpdf3d}{a,b} we plot the probability distribution function (p.d.f.) of the
 fluid velocity fluctuations ${\bm u}^f\equiv{\bm u}[c=1]$. Both the horizontal and
 vertical velocity p.d.f.'s are in quantitative agreement with the experimental data of
 \citet{risso_legendre_2010} and \citet{riss18}. The p.d.f. of the velocity fluctuations
 of the horizontal velocity components are symmetric about origin and have stretched
 exponential tails, whereas the vertical velocity fluctuations are positively skewed
 \citep{risso_legendre_2010,almeras2017,vivek_2016}. Our results are consistent with the
 recently proposed stochastic model of  \citet{risso_pdf_2016} which suggests that the
 potential flow disturbance around bubbles, bubble wakes, and the turbulent agitation
 because of flow instabilities together lead to the observed velocity distributions. We
 believe that the deviation in the tail of the distributions arises because of the
 differences in the wake flow for different $\Rey$ and $\At$ (see \fig{sing:bub}). Note
 that positive skewness in the vertical velocity has also been observed in thermal
 convection with bubbles \citep{RBLuca}.

\begin{figure}
\centering
    \includegraphics[width=0.48\linewidth]{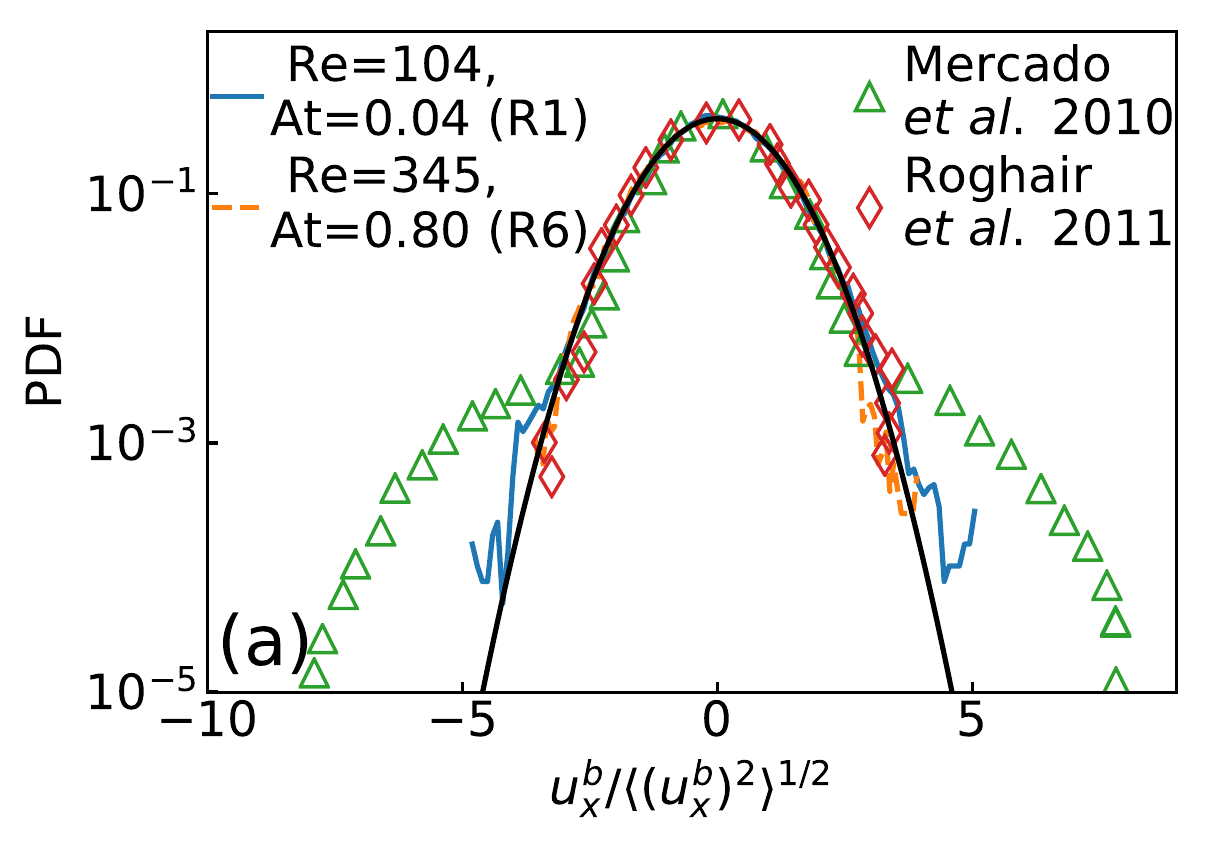}
    \includegraphics[width=0.48\linewidth]{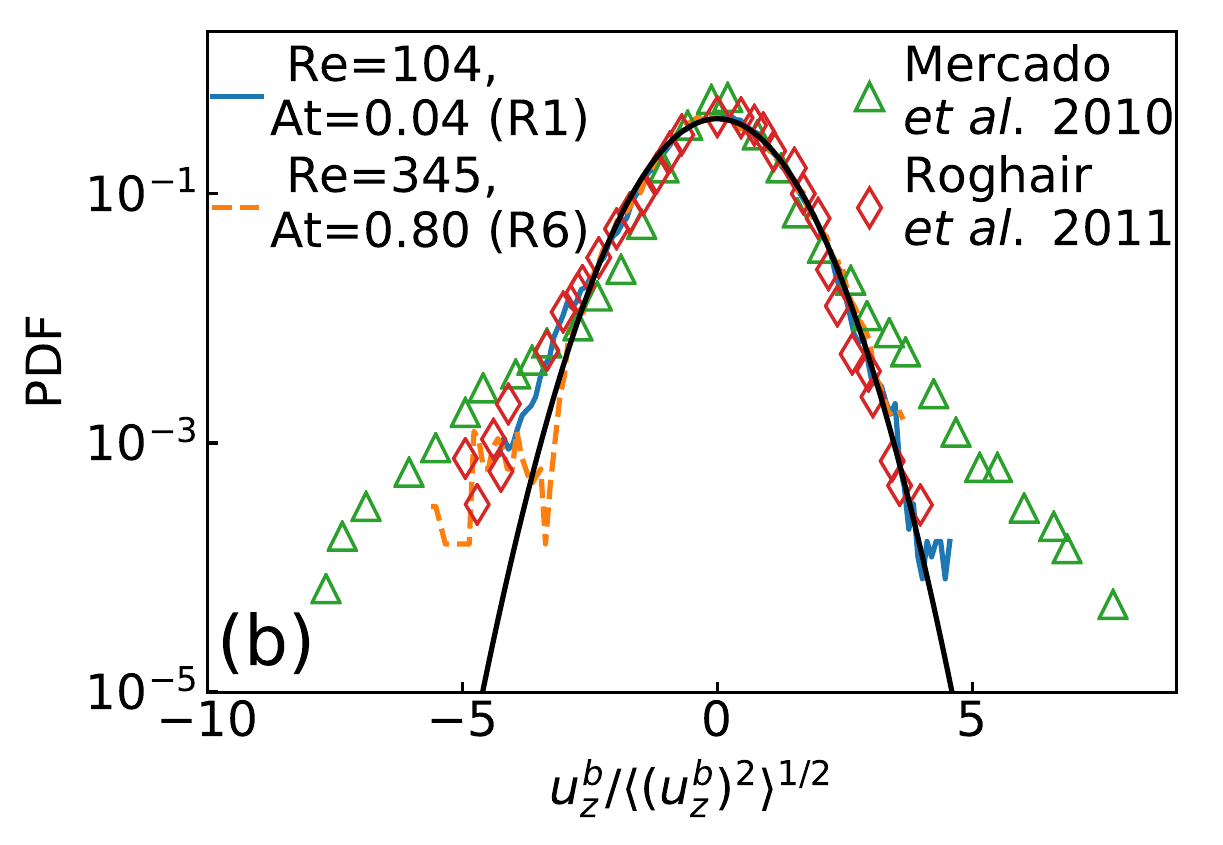}
\caption{\label{bubpdf3d} The probability distribution function of (a) the horizontal  
    and (b) the vertical component of the bubbles velocity fluctuations for runs {\tt R1}
    and {\tt R6} (see table~\ref{tab:runs}). 
    The experimental data  of \citet{martinez_2010} and numerical results of
    \citet{roghair} is also shown for comparison. The black continuous line represents a
    Gaussian distribution.
    }
\end{figure}

By tracking the individual bubble trajectories we obtain their center-of-mass velocity
${\bm u}^b$. In agreement with the earlier
 simulations of \citet{roghair}, the p.d.f.'s of the bubble velocity fluctuation are
 Gaussian (see \fig{bubpdf3d}).  The departure in the tail of the distribution is most
 probably because of the presence of large scale structures observed in experiments that
 are absent in simulations with periodic boundaries \citep{roghair}.


\subsection{Energy spectra and scale-by-scale budget}
\label{esp}

In the following, we study the energy spectrum
\begin{eqnarray}
\displaystyle E^{uu}_k &\equiv & \sum_{k-1/2<m<k+1/2} |\hat{\bm{u}}_m|^2, \nonumber
\end{eqnarray}
the co-spectrum
\begin{eqnarray}
\displaystyle E^{\rho uu}_k &\equiv & \sum_{k-1/2<m<k+1/2} \Re[\hat{(\rho {\bm u})}_{-m}
             \hat{\bm{u}}_m] \equiv d \mathscr{E}/dk, \nonumber
\end{eqnarray}
and the scale-by-scale energy budget.  Our derivation of the energy budget is similar to
\citep{frisch,pope} and does not require the flow to be homogeneous and isotropic.  For a
general field $f({\bm x})$, we define a corresponding coarse-grained field \citep{frisch} $f^<_{k}
({\bm x})\equiv \sum_{m \leq k} f_{\bm m} \exp(i {\bm m} \cdot {\bm x})$ with the
filtering length scale $\ell = 2\pi/k$. Using the above definitions in
Eq.~\eqref{eqn:mom}, we get the energy budget equation

\begin{equation}
   \partial_t\mathscr{E}_k  + \Pi_k + \mathscr{F}^\sigma_k =   \mathscr{P}_k - \mathscr{D}_k   + \mathscr{F}^g_k. 
   \label{ebud}
\end{equation}   
Here, $2\mathscr{E}_k = \langle \bm{u}^<_k\cdot(\rho\bm{u})^<_k\rangle$ is
the cumulative energy  up to wave-number $k$,  $2 \Pi_k = \langle(\rho\bm{u})^<_k
\bm{\cdot} (\bm{u}\bm{\cdot}\nabla\bm{u})^<_k\rangle +
\langle\bm{u}^<_k\bm{\cdot} (\bm{u}\bm{\cdot}\nabla\rho\bm{u})^<_k\rangle$ is the energy
flux through wave-number $k$,  $2\mathscr{D}_k=
-[\langle(\rho\bm{u})^<_k\bm{\cdot}\left(\nabla\bm{\cdot} [ 2\mu {\cal
    S}]/\rho\right)^<_k\rangle + \langle\bm{u}^<_k\bm{\cdot}(\nabla \bm{\cdot} [2 \mu
{\cal S}])^<_k\rangle]$ is the cumulative energy dissipated upto $k$, $2\mathscr{F}^\sigma_k = -[
\langle(\rho\bm{u})^<_k\bm{\cdot}\left(\bm{F}^\sigma/ \rho \right)^<_k \rangle +
\langle\bm{u}^<_k\bm{\cdot}(\bm{F}^\sigma)^<_k\rangle]$ is the cumulative energy transferred from
 the bubble surface tension to the fluid upto $k$, $2\mathscr{F}^g_k = \langle(\rho\bm{u})^<_k\bm{\cdot}\left(\bm{F}^g / \rho \right)^<_k \rangle + \langle\bm{u}^<_k\bm{\cdot}(\bm{F}^g)^<_k\rangle$ is cumulative energy injected by buoyancy upto $k$.
 In crucial departure from the uniform density flows, we find a non-zero cumulative pressure contribution $2\mathscr{P}_k = \langle (\rho\bm{u})^<_k \bm{\cdot}\left({\nabla p}/{\rho}\right)^<_k\rangle$.

In the  Boussinesq regime (small $\At$), the individual terms in the scale-by-scale budget  simplify to their uniform density
analogues: $\mathscr{E}_k=\rho_a \langle \bm{u}^<_k\cdot \bm{u}^<_k\rangle/2$,  $\Pi_k =  \rho_a \langle\bm{u}^<_k
\bm{\cdot}(\bm{u\cdot}\nabla \bm{u})^<_k\rangle$, $\mathscr{D}_k = - \mu \langle |\nabla {\bm u}_k^<|^2 \rangle$, $\mathscr{F}^\sigma_k =
-\langle\bm{u}^<_k\bm{\cdot} (\bm{F}^\sigma)^<_k\rangle$, $\mathscr{F}^g_k =\langle\bm{u}^<_k\bm{\cdot} (\bm{F}^g)^<_k\rangle$, and $\mathscr{P}_k=0$.

\subsubsection{Low $\At$  {\rm(runs} $\tt{R1}-\tt{R4}${\rm )}}

\begin{figure}
\centering
\includegraphics[width=0.45\linewidth]{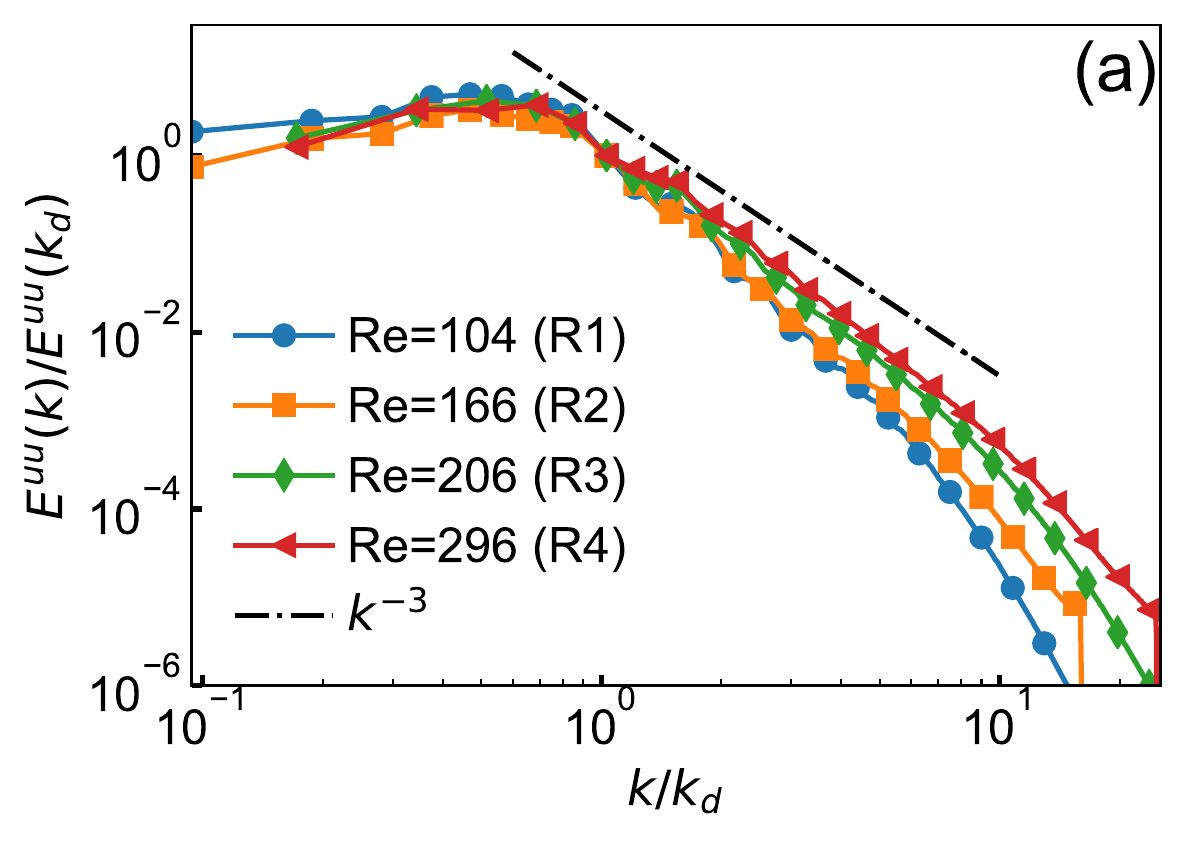}
\includegraphics[width=0.45\linewidth]{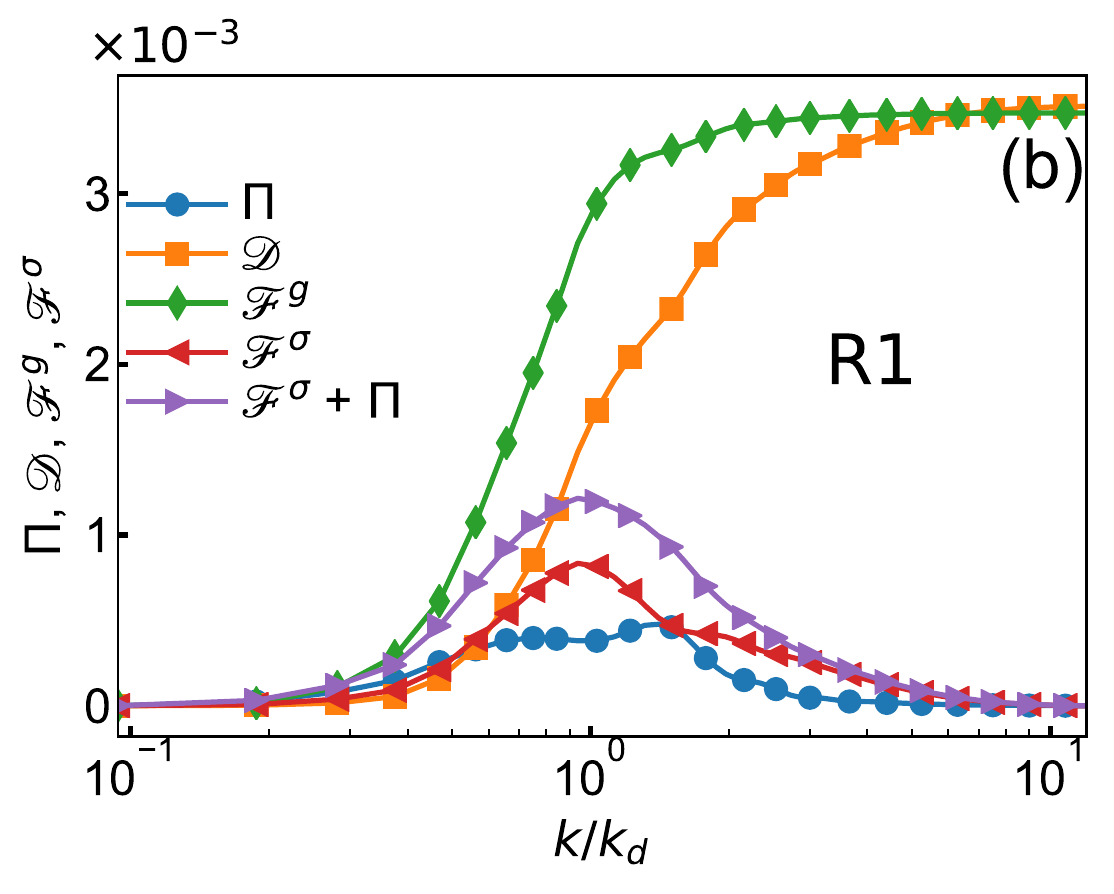}\\
\includegraphics[width=0.5\linewidth]{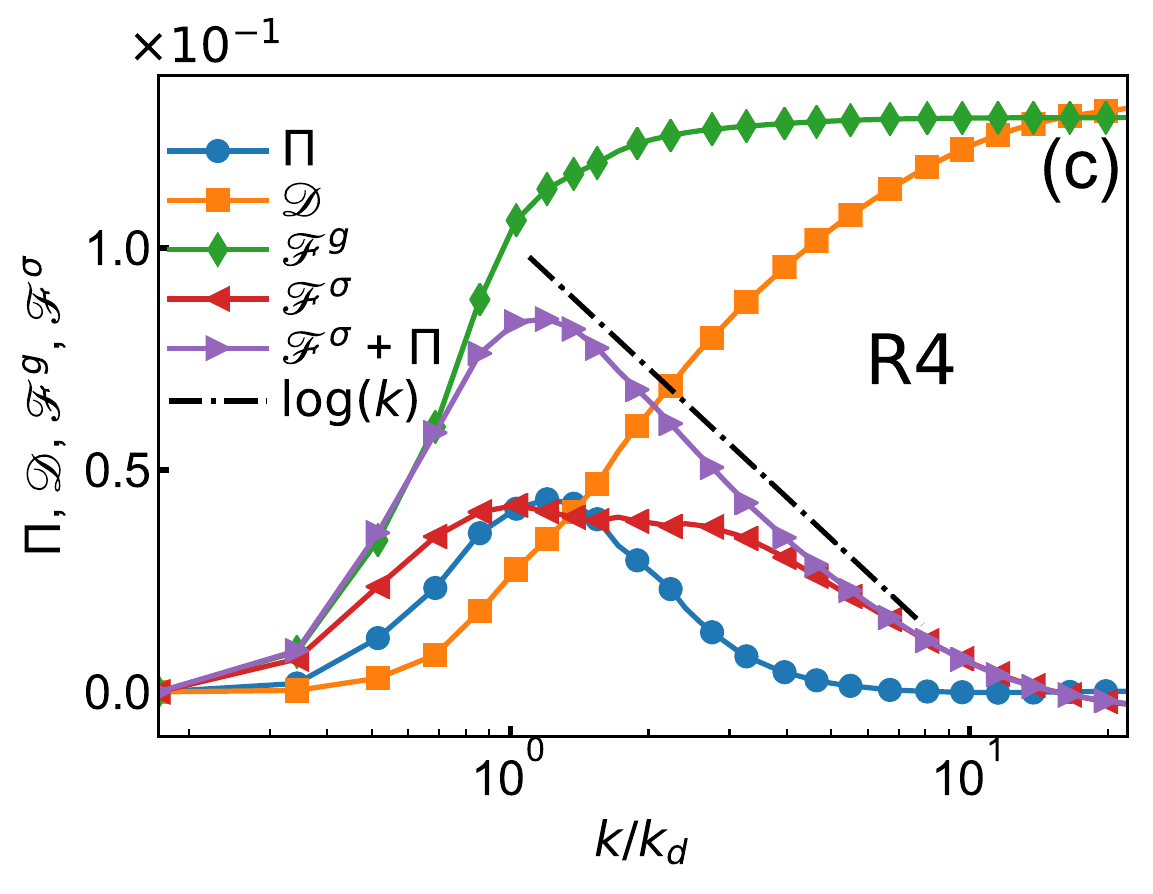}
  \caption{\label{spec:lat} (a) Log-log plot of
     energy spectra $E^{uu}_k$ versus $k/k_d$ for our high $\Rey$ low $\At$
     runs {\tt R1 - R4}. Dashed dotted line indicates the $k^{-3}$ scaling. Cumulative contribution of viscous dissipation
     $\mathscr{D}_k$, energy
     injected because of buoyancy $\mathscr{F}^g_k$ and the surface tension contribution
     $\mathscr{F}^\sigma_k$ versus $k/k_d$ for (b) run $\tt{R1}$ and (c) run {\tt R4}.
     Note that, for $k>k_d$, the balance between $d\mathscr{F}_k^\sigma/dk$ and $d\mathscr{D}_k/dk$ 
     is more prominent in panel (c) compared to (b).}
\end{figure}

We first discuss the results for the Boussinesq regime (low $\At$). For scales smaller than the bubble diameter ($k>k_d$), the energy spectrum (\subfig{spec:lat}{a})  shows a power-law behavior $E(k)\sim k^{-\beta}$ for different $\Rey$. The exponent $\beta=4$ for $\Rey=104$, it  decreases on increasing the $\Rey$ and becomes close to $\beta=3$ for the largest $\Rey=296$.   

We now investigate the dominant balances using the scale-by-scale energy budget
analysis. In the statistically steady-state $\partial_t {\mathscr E}_k=0$, and $\Pi_k +
\mathscr{F}^\sigma_k =   - \mathscr{D}_k   + \mathscr{F}^g_k$ (note that $\mathscr{P}_k=0$
for low $\At$).  In \subfig{spec:lat}{b} and \subfig{spec:lat}{c} we plot different
contributions to the cumulative energy budget for $\Rey \approx 104$ and $\Rey \approx
300$ and make the following observations:

\begin{enumerate}

\item Cumulative energy injected by buoyancy ${\mathscr F}_k^\sigma$ saturates around
  $k\approx k_d$. Thus buoyancy injects energy at scales comparable to and larger than the
  bubble diameter.
\item  Energy flux  $\Pi_k  >0 $ around $k\approx k_d$ and it vanishes for $k\gg k_d$.  
\item Especially for scales smaller than the bubble diameter, the cumulative energy
  transfer from the bubble surface tension to the velocity is the dominant energy transfer
  mechanism.
\item Consistent with the earlier predictions \citep{risso_legendre_2010}, for our highest
  $\Rey=300$ simulation provides a direct evidence that the balance of total production
  $d(\Pi_k+{\mathscr F}_k^\sigma)/dk \sim k^{-1}$  with viscous dissipation [$d{\mathscr
    D}_k/dk  = \nu k^2 E(k)$]  gives the psuedo-turbulence spectra $E(k)\sim k^{-3}$
  \citep{risso_legendre_2010,roghair,martinez_2010}.
\end{enumerate}

Our scale-by-scale analysis, therefore, suggests the following mechanism of
pseudo-turbulence. Buoyancy injects energy at scales comparable to and larger to the
bubble size. A part of the energy injected by buoyancy is absorbed in stretching and
deformation of the bubbles and another fraction is transferred via wakes to scales
comparable to bubble diameter. Similar to polymers in turbulent flows
\cite{per06,per10,valente14}, the relaxation of the bubbles leads to injection of energy
at scales smaller than the bubble diameter.

\subsubsection{High $\At$ {\rm(runs} $\tt{R5-R7}$\rm{)}}
\begin{figure}
    \centering
    \includegraphics[width=0.49\linewidth]{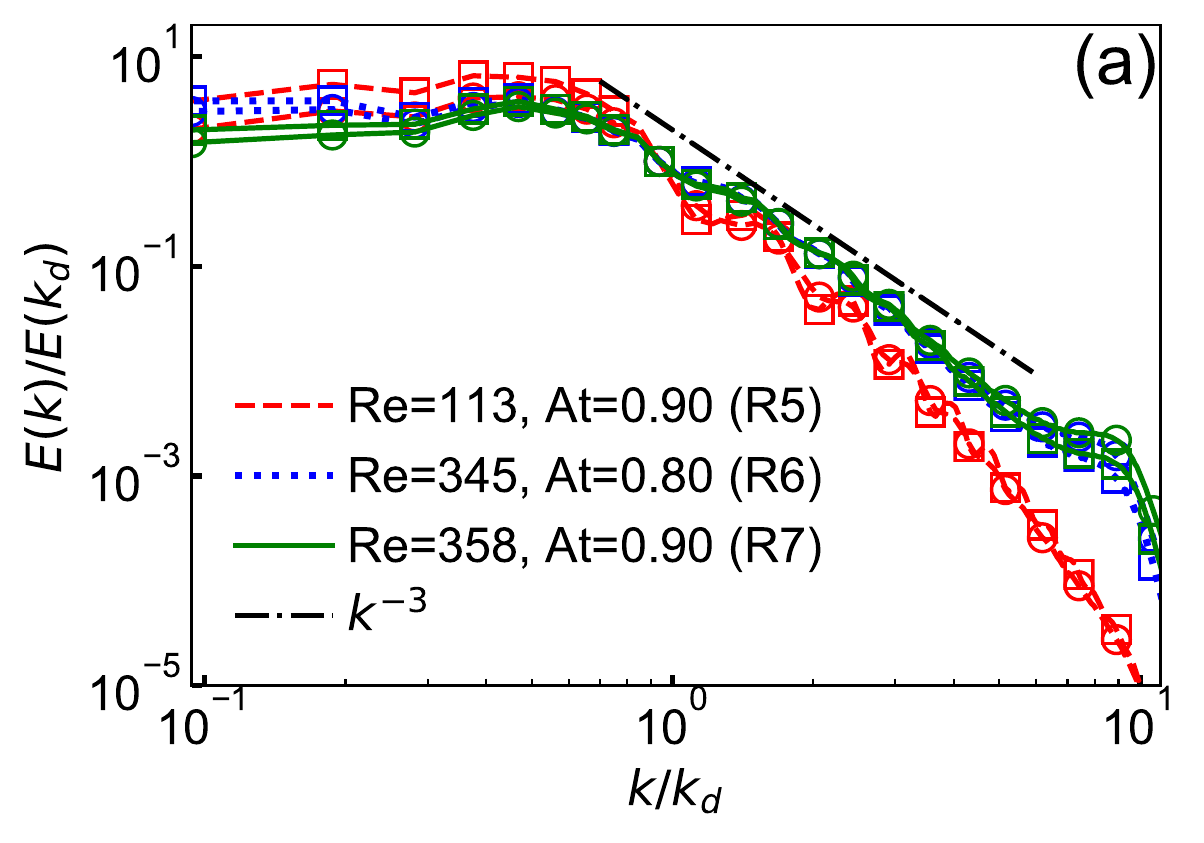}
    \includegraphics[width=0.49\linewidth]{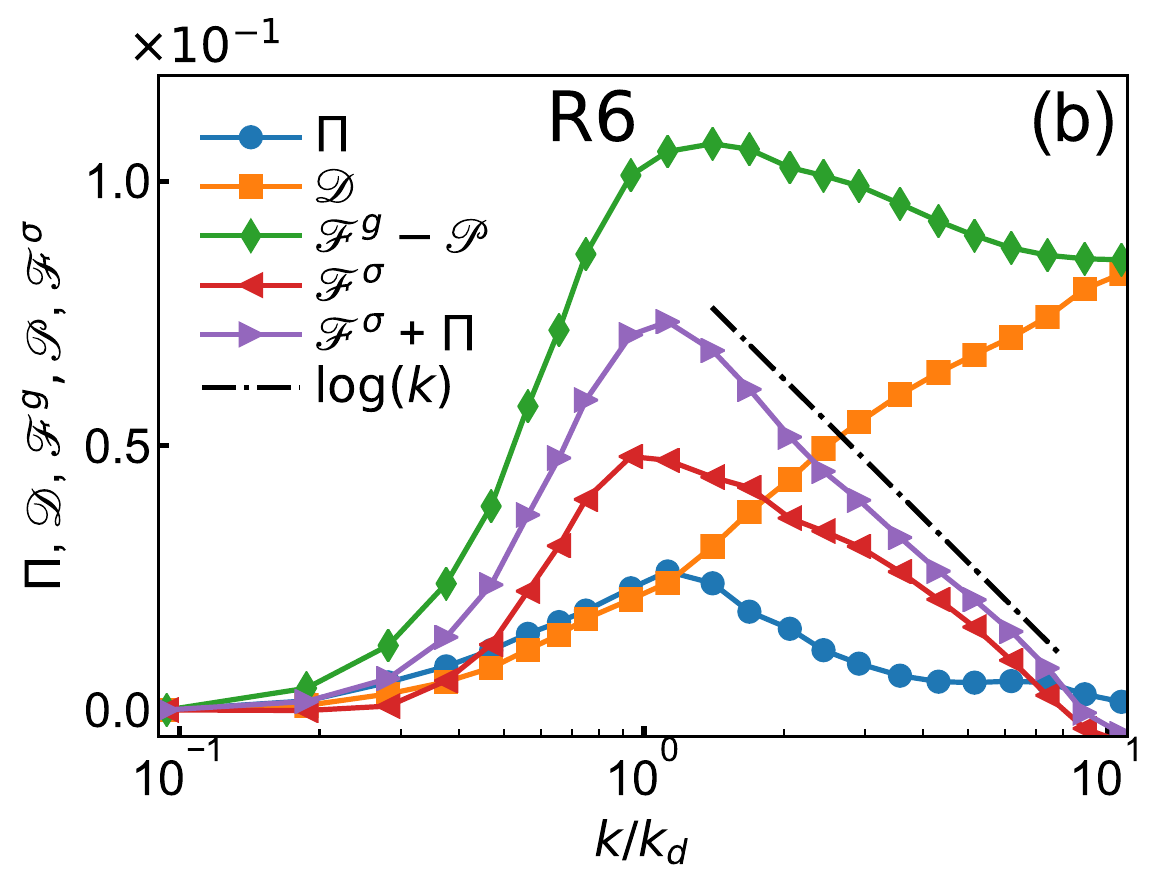}
    \caption{\label{spec:hrehat}(a) Log-log plot of
     energy spectra ($\bigcirc$) $E^{uu}_k$ and co-spectrum ($\square$) $E^{\rho uu}_k$versus $k/k_d$ for our high
     $\Rey$ high $\At$
     runs {\tt R5 - R7}. Dashed dotted line indicates the $k^{-3}$ scaling. (b) Cumulative
     contribution of the viscous dissipation $\mathscr{D}_k$, the contribution due to
     buoyancy and pressure  $\mathscr{F}_k^g-\mathscr{P}_k$, the energy flux
     $\mathcal{\Pi}_k$ and the surface tension contribution $\mathscr{F}_k^\sigma$ versus
     $k/k_d$ for run ${\tt R6}$.}
\end{figure}

Similar to earlier section, here also the energy spectrum and the co-spectrum shows a
$k^{-3}$ (\subfig{spec:hrehat}{a}). However, because of density variations the
scale-by-scale energy budget becomes more complex. Now, in the statistically steady state

 $\Pi_k + \mathscr{F}^\sigma_k =   \mathscr{P}_k - \mathscr{D}_k   + \mathscr{F}^g_k$.  

In (\subfig{spec:hrehat}{b}) we plot the scale-by-scale energy budget for our high $\At$
run ${\tt R5}$. We find that the cumulative energy injected by buoyancy and the pressure
contribution ${\mathscr F}_k^g + \mathscr{P}_k$  reaches  a peak around $k \approx k_d$
and then decrease mildly to $\epsilon_{inj}$. Similar to the low $\At$ case, we find a
non-zero energy flux for $k\approx k_d$ and a dominant surface-tension contribution to the
energy budget for $k\gg k_d$. Finally, similar to last section, for $k>k_d$ the net
production $d (\Pi + \mathscr{F}^\sigma)/dk \sim k^{-1}$ balances viscous dissipation $\nu
k^2 E(k)$ to give $E(k)\sim k^{-3}$.

\section{Conclusion}
\label{concl}
To conclude, we have investigated the statistical  properties of velocity fluctuations in psuedo-turbulence 
generated by buoyancy driven bubbly flows.  The $\Rey$ that we have explored are
consistent with the $\Rey \sim[300-1000]$ used in the experiments
\citep{risso_legendre_2010,vivek_2016,mendez}.  Our numerical simulations show that the shape of the p.d.f. of
the velocity fluctuations is consistent with experiments over a wide range of $\Rey$ and
$\At$ numbers. For large $\Rey$ and for low as well as high $\At$, the energy spectrum
shows a $k^{-3}$ scaling but it becomes steeper on reducing the $\Rey$.   We observe a
non-zero positive energy flux  for scales comparable to the bubble diameter.  Our
scale-by-scale energy budget validates the theoretical prediction that the net production
balances viscous dissipation to give $E(k)\sim k^{-3}$.

\section{Acknowledgments}
We thank D. Mitra and S. Banerjee for discussions. This work was supported by research
grant No. ECR/2018/001135 from SERB, DST (India).


\end{document}